\documentclass[5p,twocolumn]{elsarticle}
\usepackage{graphicx,latexsym}
\usepackage{dcolumn}
\usepackage{amssymb,amsmath,bm}
\usepackage{subfigure}
\usepackage{braket}
\usepackage{siunitx}
\usepackage{array}
\usepackage{float}
\usepackage[nopar]{lipsum}
\usepackage{caption}
\usepackage{multirow}
\usepackage{bigstrut}
\usepackage[super]{nth}

\usepackage{hyperref}
\hypersetup{
    pdfnewwindow=true,       % links in new window
    colorlinks=true,         % false: boxed links; true: colored links
    linkcolor=blue,          % color of internal links
    citecolor=blue,          % color of links to bibliography
    filecolor=magenta,       % color of file links
    urlcolor=black           % color of external links
}

\usepackage[normalem]{ulem}

\def\sec#1{Sec.\ \ref{#1}}
\def\eq#1{Eq.\ (\ref{#1})}
\def\fig#1{Fig.\ \ref{#1}}

\journal{}

\begin{document}

\begin{frontmatter}

%-----------------------------------------------------------------

\title{Photon and magnetic field controlled electron transport of a multiply-resonant\break
	   photon-cavity double quantum dot system}

\author[a1,a2]{Halo Anwar Abdulkhalaq}
\ead{halo.abdulkhalaq@univsul.edu.iq}
\address[a1]{Division of Computational Nanoscience, Physics Department, College of Science,
	University of Sulaimani, Sulaimani 46001, Kurdistan Region, Iraq}
\address[a2]{Computer Engineering Department, College of Engineering, Komar University of Science and Technology, Sulaimani 46001, Kurdistan Region, Iraq}

\author[a1,a2]{Nzar Rauf Abdullah}
\ead{nzar.r.abdullah@gmail.com}

\author[a4]{Vidar Gudmundsson}
\address[a4]{Science Institute, University of Iceland, Dunhaga 3, IS-107 Reykjavik, Iceland}
\ead{vidar@hi.is}

%----------------------------------------------------------------

\begin{abstract}
We study electron transport through double quantum dots (DQD) coupled to a cavity with a single photon mode. The DQD is connected to two electron reservoirs, and the total system is under an external perpendicular magnetic field. The DQD system exhibits a complex multi-level energy spectrum. By varying the photon energy, several anti-crossings between photon dressed electron states of the DQD-cavity system are found at low strength of the magnetic field. The anti-crossings are identified as multiple Rabi resonances arising from the photon exchange between these states. As the results, a dip in the current is seen caused by the multiple Rabi resonances. By increasing the strength of the external magnetic field, a dislocation of the current dip to a lower photon energy is found and the current dip can be diminished.
The interplay of the strength of the magnetic field and the geometry of the states the DQD system
can weaken the multiple Rabi resonances in which the exchange of photon between the anti-crossings is decreased. We can therefore confirm that the electron transport behavior in the DQD-cavity system can be controlled by manipulating the external magnetic field and the photon cavity parameters.

\end{abstract}


\end{frontmatter}

\section{Introduction}

Quantum dots (QDs) are proving to be a very promising solution for a range of electronic and optical applications in new technologies, due to their outstanding properties \cite{doi:10.1021/acsanm.0c01386}.
The open quantum system consisting of quantum dots connected to electron reservoirs is a good framework for investigating electronic and optical properties of QDs including electron transport \cite{RevModPhys.75.1, ABDULLAH2020114221}.
Controlling the electron transport through a QD or a double quantum dot (DQD) could improve many applications in several different fields \cite{2013}. Few of the important applications are biological imaging \cite{jin2011application}, quantum computing \cite{nakamura1999coherent}, quantum dots battery \cite{bagher2016quantum} and nanowire photodetectors \cite{van2010single}.

Due to the charge quantization effect, the electron transport through the DQD occurs owing to the Coulomb interaction effect which maintains the transport in the Coulomb blockade (CB) regime \cite{Stephanie,averin1991theory}. Many aspects of the electron transport under the Coulomb interaction have enriched the interest in this area, such as the study of the CB in atomically thin QD \cite{brotons2019coulomb}, transport in multilevel QDs \cite{yeyati1999transport}, and magnetoresistance in the CB regime \cite{takahashi1998effect, ono1996magnetoresistance}. Moreover, the shuttle mechanism in the charge transfer \cite{gorelik1998shuttle}, the transient current spectroscopy of a quantum dot \cite{fujisawa2001transient}, the CB oscillations in the thermopower of a quantum dot \cite{staring1993coulomb}, and the conductance of a proximitized nanowire \cite{van2016conductance} are among the conducted studies.

The transport in the interacting Coulomb system can be controlled by a photon source in which the so called photon-assisted transport (PAT) has been reported \cite{PhysRevB.50.2019, 2019}. In this way of controlling the electron motion the photon frequency of the source radiation should match that of the electronic system \cite{ABDULLAH2014254}, this makes the electron transport to be enhanced or reduced by the photon emission and absorption processes \cite{torres2005mono}. The PAT is classical if it is induced by a classical electromagnetic field, while the electron transport can also be influenced by a single photon, a quantized photon field, hence, inducing a quantum PAT \cite{PhysRevLett.73.3443}.
The DQD coupled to a photon cavity have been studied and it has been shown that  DQD-cavity systems can be realized experimentally using an array of superconducting quantum interference devices \cite{PhysRevX.7.011030}. In such a DQD-cavity, a current suppression is seen when the ac-field forms spin-blocked states \cite{PhysRevB.100.195421}. Furthermore, if the electron-phonon interaction is taken into account in the DQD system, the amplitude and the phase response of the cavity field exhibit oscillations that are periodic in the DQD energy level with a detuning due to the phonon modes \cite{PhysRevLett.120.097701}.
It has been recently shown that an absorbed photon gives rise to a single electron tunneling through a double dot, with a conversion efficiency reaching $6\%$ \cite{Khan2021}. The aforementioned results may have potential applications in the fields of quantum optics and quantum information science \cite{Xing_2021}.

There are several approaches employed to investigate the phenomena of quantum transport in DQD systems coupled to a photon source. One of the techniques to study the electron transport in QD systems is applying non-equilibrium Green's functions, which can provide a conceptual basis to develop quantitative models of quantum transport \cite{Greendatta2002non}. This technique has also been used in the calculation of a transmission function of a QD system that is coupled to a photon field and side-coupled to a topological superconductor nanowire embedded with double Majorana bound states (MBSs) \cite{10.3389/fphy.2020.00254}. Another approach is the quantum master equation (QME), which is used to derive the many body density matrix of a weakly coupled current-carrying open system to two metal leads \cite{harbola2006quantum}. The QME is used to derive a quantum rate equation to describe the resonant transport in mesoscopic systems as shown in \cite{gurvitz1996microscopic}. In a DQD system different approaches of master equations have been used to investigate the role of coherence in thermoelectric transport \cite{pirot2022thermal}.

The system of our study, which is a finite quantum wire embedded with double quantum dots (DQD), is connected to two semi-infinite leads having different chemical potential values. The system is coupled to a single photon cavity, and we will show how this affects the electron transport. The total system is placed in a static external magnetic field that also affects the electron transport in the DQD. The presence of a magnetic field affects the motion of the electrons leading to edge and formation of localized states, which results in reduction of the electron transport \cite{abdullah2016competition, abdullah2016effects}.
We study the properties of the electron transport in the steady-state regime in a strong coupling regime ($g_{\rm ph} > \kappa$), where the $g_{\rm ph}$ and $\kappa$ are the electron-photon and the cavity-photon field to environment coupling strengths, respectively.

The organization of the paper will be as follows: In \sec{Model_Theory}.\ we describe the model and the theoretical formalism used. The obtained results are shown in \sec{Results}, and in \sec{Conclusion}. conclusions will be presented.

\section{The Model and Theoretical Formalism}\label{Model_Theory}

In this section, we display the Hamiltonians describing the parts of the system and the theoretical formalism of the master equation describing the open central system under the influence of the environment. We consider a double quantum dot (DQD) which is embedded in a quantum wire. The DQD system is made of GaAs material with the effective mass of $m^* = 0.067 m_{\rm e}$. The DQD system is coupled to a photon cavity and connected to two electron reservoirs from both ends. The quantum wire is a two dimensional system which is hard-wall confined in the $x$-direction at $x=\pm L_x/2$ and parabolically confined in the $y$-direction, where $L_{x}$ is length of the quantum wire in the $x$-direction.

We start with the DQD potential of our system and the confinement parameters.
The system of our study is a parabolic quantum wire embedded with two symmetrical quantum dots placed
asymmetrically, is described by the potential \cite{2019b}
\begin{equation}
\begin{split}
	V(x,y)
	&=\bigg[\frac{1}{2}m^*\Omega_0^2y^2-eV_g+\theta\bigg(\frac{L_x}{2}-|x|\bigg)\times V_{\rm DQD}\bigg]
\end{split}
\end{equation}
with
\begin{equation}
	V_{\rm DQD}=\sum_{i=1}^{2}V_{d}^i
	\exp{\{-\beta_i^2(x-x_{0i})^2-\beta_i^2(y-y_{0i})^2\}},
\end{equation}
where the values of the parameters are as follows; $m^*$ is electron effective mass, the bare confinement energy $\hbar\Omega_0=2.0$ meV, $e$ is electron charge and $V_g=0.6$ meV is plunger gate voltage. $\theta$ is the Heaviside unit step function and $L_x=150$ nm is the length of the wire.
The depth of both the first and the second quantum dot is $V_d^1=V_d^2=-4.5$ meV, $\beta_1=\beta_2= 0.03$ nm$^{-1}$, $x_{01}=23.802$ nm, $x_{02}=-23.802$ nm, $y_{01}=47.604$ nm, $y_{02}=-47.604$ nm. Here, the plunger gate voltage, $V_g$, shifts the electron energy states into or out of the bias window which is set by the left and right leads chemical potentials.

\begin{figure}[htb]
	\centering
	\includegraphics[width=0.35\textwidth]{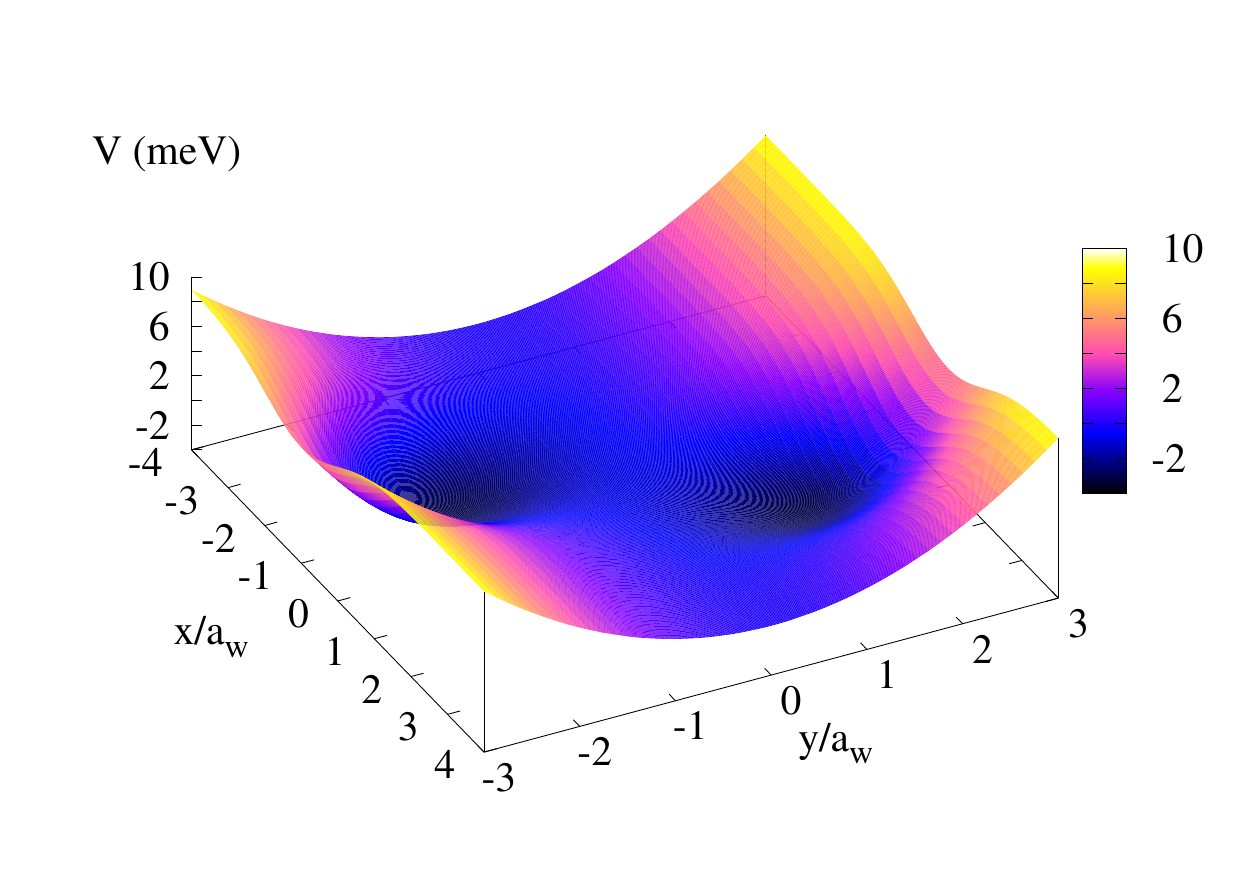}
	\caption{The potential representing a parabolic quantum wire embedded with two quantum dots.
	$B = 0.1$ T, $a_w=23.802$ nm, and length of the quantum wire $L_x=150$ nm.}
	\label{fig01}
\end{figure}

The total system, the DQD system and the leads, is initially placed in a weak perpendicular external magnetic field $\textbf{B}=B\hat{\bm z}$ with magnitude of $0.1$ T. The effective confinement frequency due to the magnetic field is given by $\Omega_w^2=w_c^2+\Omega_0^2$, where the $w_c=eB/m^*c$ is the cyclotron frequency. To scale the system the effective magnetic length, $a_{\omega}=(\hbar/m^*\Omega_{\omega})^{1/2}$, is used as shown in \fig{fig01}.
Later on, we vary slightly the strength of the external magnetic field.

A many-body (MB) Hamiltonian, $H_{\rm S}$, (of the DQD system, and the cavity) describes the electrons in the central system controlled by a plunger gate voltage and strongly coupled to a single mode photon field. The $H_{\rm S}$ explicitly contains the electron-electron interaction and the electron-photon coupling. Hence, the Hamiltonian of the total system is defined as
%%%
\begin{equation}
	H_{\rm S} = H_{\rm e} + H_{\rm ph} + H_{\rm e\text{-}ph},
	\label{H_S}
\end{equation}
%%%
where $H_{\rm e}$ is the DQD system Hamiltonian in a short quantum wire, $H_{\rm ph}$ is the free photon field Hamiltonian, and $H_{\rm e\text{-}ph}$ is the DQD system to the photon cavity coupling Hamiltonian. Starting with the DQD system Hamiltonian that is written as \cite{2019}
%%%
\begin{equation}
	\begin{split}
		H_{\rm e} & =\sum_{nn'}\bra{\Psi_{n'}}\left[\frac{\pi_e^2}{2m^*}+eV_g+V_{\rm DQD}\right]\ket{\Psi_n}d_{n'}^{\dagger}d_n \\
		& +H_Z+\frac{1}{2}\sum_{nn'mm'}V_{nn'mm'}d_{n'}^{\dagger}d_{m'}^{\dagger}d_nd_m,
	\end{split}
\label{H_e}
\end{equation}
%%%
where $|\Psi_n\rangle$ is a single-electron eigenstate (SES), and $d_{n'}^{\dagger}$ and $d_n$ are the central system electron creation and annihilation operators, respectively. $\pi_e=\textbf{p}+\frac{e}{c}\textbf{A}_{ext}$, where $\textbf{p}$ is the momentum operator and $\textbf{A}_{ext}=-B_y\hat{x}$ is the vector potential of the external magnetic field \textbf{B}. The second term of \eq{H_e}, $H_Z = (\mu_B \, g_{\rm S} \, B / 2) \sigma_z$, is the Zeeman Hamiltonian, where $\mu_B$ is the Bohr magneton, $g_{\rm S}$ is the electron spin $g$-factor, $\sigma_z$ is a Pauli-spin matrix. The third term represents the Coulomb interaction containing the Coulomb integral
%%%
\begin{equation}
	V_{nn'mm'}=\bra{\Psi_{n'}\Psi_{m'}}\left[\frac{e^2}{\bar{\kappa}|r-r'|}\right]\ket{\Psi_n\Psi_m},
\label{Vcoul}
\end{equation}
%%%
where $\bar{\kappa}$ is dielectric constant, and $|r-r'|$ is an electron pair spatial separation.
The free photon field Hamiltonian, $H_{\rm ph}$, in \eq{H_S} is given by
%%%
\begin{equation}
	H_{\rm ph} = \hbar \omega_{\rm ph} N_{\rm ph},
\end{equation}
%%%
where $\hbar \omega_{\rm ph}$ is the photon energy, and $N_{\rm ph} = a^{\dagger}a$ is photon number operator with $a^{\dagger}$ and $a$ being, respectively, the photon creation and annihilation operators. The third term in \eq{H_S} is the electron-photon interaction Hamiltonian given by
%%%
\begin{equation}
	\begin{split}
		H_{\text e\text{-}ph} & =\frac{e}{m^*c} \sum_{nn'}\bra{\Psi_{n'}}{\bm \pi}_e\cdot\textbf{A}_{ph}\ket{\Psi_n}d_{n'}^{\dagger}d_n \\
		& +\frac{e^2\textbf{A}_{\rm ph}^2}{2m^*c^2}\sum_{nn'}\braket{\Psi_{n'}|\Psi_n}d_{n'}^{\dagger}d_n,
	\end{split}
\label{Heg}
\end{equation}
%%%
where $\textbf{A}_{\rm ph}$ is the quantized vector potential of the photon field given by
%%%
\begin{equation}
	\textbf{A}_{\rm ph}=A(a+a^{\dagger}) \hat{\textbf{e}},
\end{equation}
%%%
with $A$ the amplitude of the photon field, $\hat{\textbf{e}}=(e_x,0)$ and $\hat{\textbf{e}}=(0,e_y)$ are the unit vectors defining the $x$-polarization and $y$-polarization of the photon field, respectively. From the vector potential amplitude we can define the electron-photon coupling strength as $g_{\rm ph}= A \Omega_w a_w/c$. The wavelength of the FIR (far-infrared) cavity field is much larger than the size of the DQD system. This will allow to use numerical exact diagonalising technique, using a tensor product of the Coulomb interacting MB bases and the photon number operator for the electron-photon Hamiltonian \cite{2019b,442012}.

As mentioned earlier, the two leads with different chemical potentials are coupled to the DQD system from left and right side. This makes electrons to flow from the leads to the DQD system and also oppositely \cite{2019}. The leads are defined via the Hamiltonian
%%%
\begin{equation}
	H_l=\int d\textbf{q}\epsilon_l(\textbf{q})c^{\dagger}_{\textbf{q}l}c_{\textbf{q}l},
\end{equation}
%%%
where $\epsilon_l$ is the single-electron subband energy of the lead $l$, $\textbf{q}$ is the combined momentum and band index quantum number, and $c^{\dagger}_{\textbf{q}l}$ and $c_{\textbf{q}l}$ are the lead $l$ electron creation and annihilation operators, respectively \cite{ABDULLAH2014254}.\\
The Hamiltonian describing the system-lead coupling is given as
%%%
\begin{equation}
	H_{Tl}=\sum_{n}\int d\textbf{q} \left[ c^{\dagger}_{\textbf{q}l}T_{\textbf{q}nl}d_n+d^{\dagger}_n(T_{\textbf{q}nl})^*c_{\textbf{q}l}\right],
\end{equation}
%%%
where $T_{\textbf{q}nl}$ are the state-dependent coupling coefficients describing the transfer of an electron between the central system SE state $\ket{n}$ and the leads extended state $\ket{\textbf{q}}$, given by
%%%
\begin{equation}
	T_{\textbf{q}nl}=\int d\textbf{r}d\textbf{r}'\Psi_{\textbf{q}l}(\textbf{r}')^*g_{\textbf{q}nl}(\textbf{r},\textbf{r}')\Psi_n^S,
\end{equation}
%%%
where $\Psi_{\textbf{q}l}$ and $\Psi_n^S$ are the SE wavefunctions in the leads and DQD system respectively, and $g_{\textbf{q}nl}$ is the coupling function defining the contact areas \cite{gudmundsson2009time}.

The index $n$ in Eq.\ \ref{H_e} is a simple natural integer labeling the exact single-electron states of
the closed central system used to form the MB Fock space for the electrons. The single-
electrons states (SES) are 36 and the Fock space includes 1228 many-electron states (MES)
(one-, two-, and three-electron states to cover the energy range well above the bias window).
The MES are used
to diagonalize the Coulomb interaction (\ref{Vcoul}). The electron-photon interaction (\ref{Heg})
is diagonalized
using 512 of the Coulomb interacting MES tensor multiplied by the 17 lowest eigenstates
of the photon number operator. The 128, lowest in energy, fully interacting photon dressed electron states are used for the transport calculations.

The electron transport through the DQD system coupled to the leads in the steady state regime can be calculated using the Markovian master equation. This equation is based on the dynamical projection of the total system onto the DQD system, which is derived from the Nakajima and Zwanzig generalized master equation (GME) \cite{nakajima1958quantum, zwanzig1960ensemble}. The kernel of the Nakajima Zwanzig integro-differential equation for the reduced density operator for the central system is approximated
to include up to second order terms in the system leads coupling \cite{2019a}. In the steady state regime, as it is of our interest, a Markovian approximation for the GME is used transforming it to the Liouville space of transitions \cite{342017}. The reduced density operator $\rho_{\rm S}$, which describes the time evolution of the central system under the influence of the photon reservoir and the external leads, is calculated by Markovian master equation \cite{Vidar_2021} as is presented in \cite{45b2018}. The total density operator $\rho(t_0)$, before the coupling of central system to
the external leads is assumed to be a tensor product of the uncorrelated central system and external leads reduced density operators $\rho_l(t_0)$ and $\rho_{\rm S}(t_0)$, respectively
%%%
\begin{equation}
	\rho(t_0)=\rho_l(t_0)\rho_{\rm S}(t_0),
\end{equation}
%%%
where $t_0$ represents a time before the coupling. Then at $t>t_0$, after the coupling, the central system reduced density operator is written as \cite{342017}
%%%
\begin{equation}
	\rho_{\rm S}(t)=Tr_l(\rho),
\end{equation}
%%%
where $l$ represents the electron reservoirs, the left L and the right R leads.
The current passing through the DQD system, after obtaining the reduced density operator, can be calculated by
%%%
\begin{equation}
	I_{L,R}=Tr[\rho_{\rm S}^{L,R}(t) \, Q]
\end{equation}
%%%
where $\rho_S^{L,R}(t)$ is reduced density operator, and $Q=-e\sum_{n}d_n^{\dagger}d_n$ is the DQD system charge operator \cite{2019}.

\section{Results}\label{Results}

\begin{figure}[htb]
	\centering
	\includegraphics[width=0.45\textwidth]{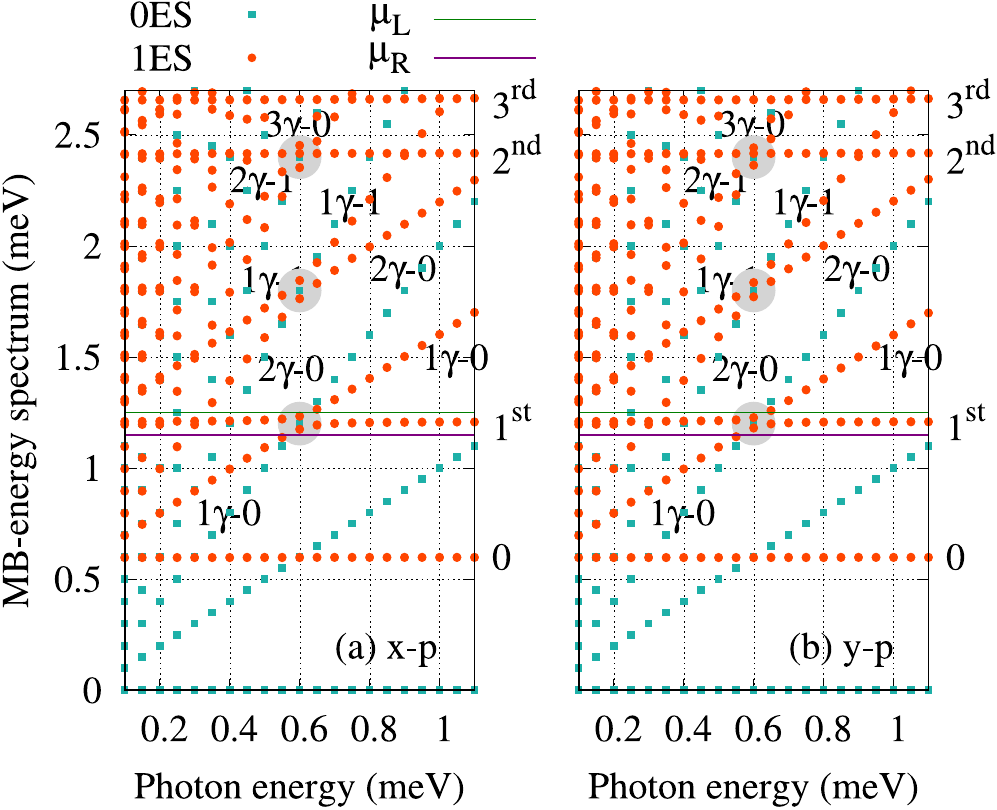}
	\caption{Many-Body (MB) energy spectrum of the DQD system coupled to photon cavity as a function of the photon energy for s photon field with (a) $x$- or (b) $y$-polarization, where 0ES (green squares) represent zero-electron states, and 1ES (red circles) are one-electron states. The green and purple lines are the chemical potential of left lead, $\mu_L=1.25$ meV, and right lead, $\mu_R=1.15$ meV, respectively. 0 is one-electron ground-state energy, while $1^{\rm st}$, $2^{\rm nd}$, and $3^{\rm rd}$ are one-electron first-, second-, and third-excited state respectively. Also, $1\gamma\text{-}0$, $2\gamma\text{-}0$, and $3\gamma\text{-}0$, respectively, refer to one-, two-, and three-photon replica of ground state, while $1\gamma\text{-}1$ and $2\gamma\text{-}1$ are one- and two-photon replica of excited states respectively. The initial photon number in the cavity is $n_{\text{R}}=1$, $g_{\rm ph}=0.1$ meV and  $\kappa=10^{-5}$ meV. The external magnetic field is $B=0.1$ T, $eV_{\rm g}=0.6$ meV, and  $T_{L,R}=0.5$ K.}
	\label{fig02}
\end{figure}

In this section, we present our results. We present the effects of varying photon energy, $\hbar\omega_{\rm ph}$, and the external magnetic field strength, on the properties of the electron transport through the DQD system.
In \fig{fig02}, the MB-energy spectrum of the DQD system coupled to cavity as a function of the photon energy is shown, where the initial photon number in the cavity is $n_{\text{R}}=1$, $g_{\rm ph}=0.1$ meV, and $\kappa=10^{-5}$ meV. The green and purple lines are the chemical potentials of left and right lead, respectively.
The energy states are categorized such that, as shown, the 0ES (green squares) are zero-electron states, and 1ES (red circles) are one-electron states. We have selected an energy range such that the four lowest energy states including the one electron ground state ($0$) the one-electron first- ($1^{\rm st}$), second- ($2^{\rm nd}$), and third-excited state ($3^{\rm rd}$) in the interval from $0.0$ to $2.7$~meV. These states have almost flat dispersions in the energy spectrum, and one can see that the $1^{\rm st}$ is located in the bias window throughout all the photon energy.
In addition to these four states, the photon replica states generated due to the photon cavity are seen in the energy spectrum. These replica states include $1\gamma\text{-}0$, $2\gamma\text{-}0$, and $3\gamma\text{-}0$, respectively, referring to one-, two-, and three-photon replica of ground state, while $1\gamma\text{-}1$ and $2\gamma\text{-}1$ are one- and two-photon replica of excited states respectively. The location of the photon replica states in the energy spectrum depends on the photon energy giving rise to a strong dispersion of these states. As we use exact diagonalization and not
simple perturbation theory, we not that more appropriately we should talk about the first, the second,
and higher order photon replicas as the mean photon content of these states is not necessarily an
integer.

It can be seen that there is no noticeable difference between the (a) $x$- and (b) $y$-polarization of the photon field to the MB-energy spectrum. This is due to the symmetric behavior of the quantum dots in which the photon field has symmetric influence on the DQD in both $x$- and $y$-direction. This can be confirmed by a single quantum dot coupled to a photon cavity in which a noticeable change in the energy spectrum of a single dot was seen for an $x$- and a $y$-polarization of a photon field \cite{2019}.
The Zeeman splitting of the external magnetic field simplifies the identification of the spin
component of each electron state.

It can be seen that the tuning of the photon energy creates some anti-crossings in the energy spectrum. There are three significant anti-crossing formed at photon energy $\hbar\omega_{\rm ph}=0.6$ meV, such that the lowest energy crossing is between the $1^{\rm st}$ and $1\gamma\text{-}0$, the next is between $2\gamma\text{-}0$ and $1\gamma\text{-}1$, and the third one is between $3\gamma\text{-}0$ and $2\gamma\text{-}1$. These anti-crossings are highlighted with gray circles in the energy spectra. The anti-crossings can play an important role in controlling electron transport as it will be shown later.

\begin{figure}[htb]
	\centering
	\includegraphics[width=0.45\textwidth]{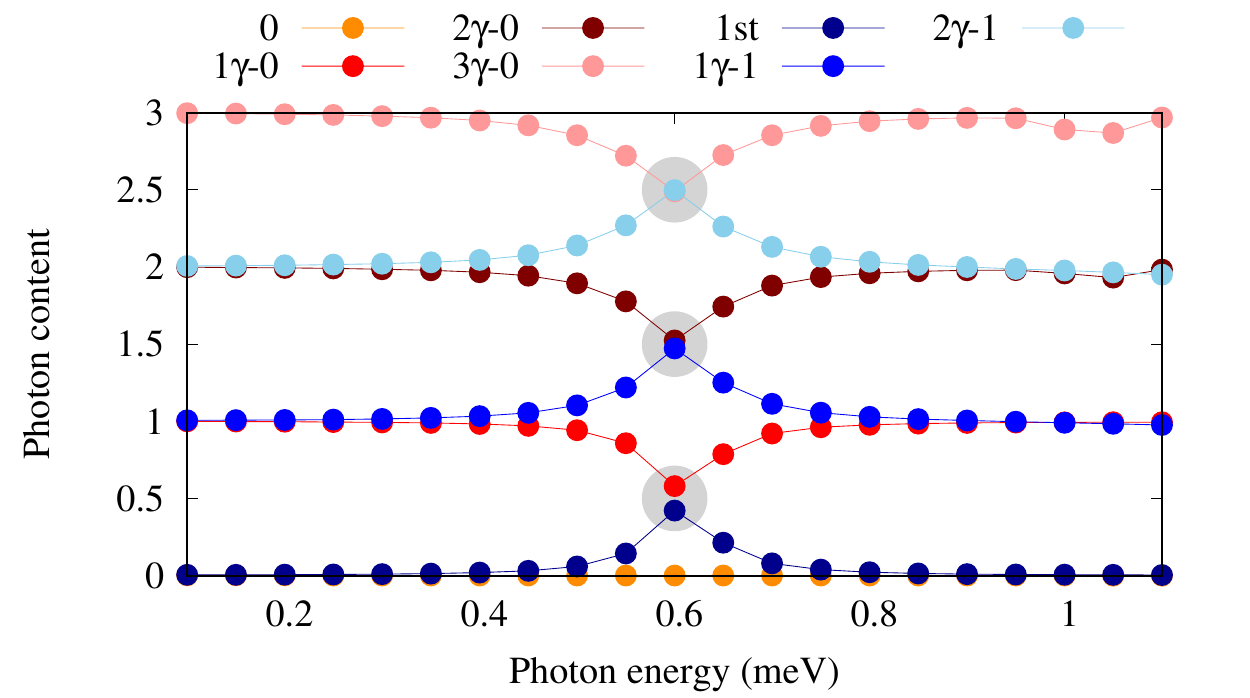}
	\caption{The photon content of the selected states as a function of the photon energy for the DQD system coupled to a cavity photon field, where the photon content is independent of the polarization of the photon field. 0 is the one-electron ground-state energy (orange), $1\gamma\text{-}0$ (red), $2\gamma\text{-}0$(maroon), and $3\gamma\text{-}0$ (pink), respectively, refer to the first-, the second-, and the third-photon replica of the ground state. While $1^{st}$ (dark blue) is the one-electron first-excited state, $1\gamma\text{-}1$ (blue) and $2\gamma\text{-}1$ (cyan) are the first- and the second-photon replica states of the first-excited state, respectively. The initial photon number in the cavity is $n_{\text{R}}=1$, $g_{\rm ph}=0.1$ meV and $\kappa=10^{-5}$ meV. The external magnetic field is $B=0.1$ T, $eV_{\rm g}=0.6$ meV, and $T_{L,R}=0.5$ K.}
	\label{fig03}
\end{figure}

These anti-crossings are caused by Rabi resonances between the states. This can be confirmed by the exchange of photons between these states, as is presented in \fig{fig03}.
The photon exchange between the anti-crossing states occurs due to the resonances between these states. The photon exchange occurs between $1^{\rm st}$ (dark blue) and $1\gamma\text{-}0$ (red), $2\gamma\text{-}0$ (blue) and $1\gamma\text{-}1$ (maroon), and $3\gamma\text{-}0$ (pink) and $2\gamma\text{-}1$ (light blue). It can be clearly seen that the photon replicas of the ground state loose their photons, while the first-excited state and its replicas gain photons in the photon energy range from $0.4$ to $0.7$ meV.
These multiple Rabi resonance states will influence the electron occupation of the DQD system and thus the electron transport. We see that the photon polarization does not much influence the energy spectrum of the DQD system, and will also not have much effect on the transport properties as well. We therefore present only the results for $x$-polarization below.

\begin{figure}[htb]
	\centering
	\includegraphics[width=0.45\textwidth]{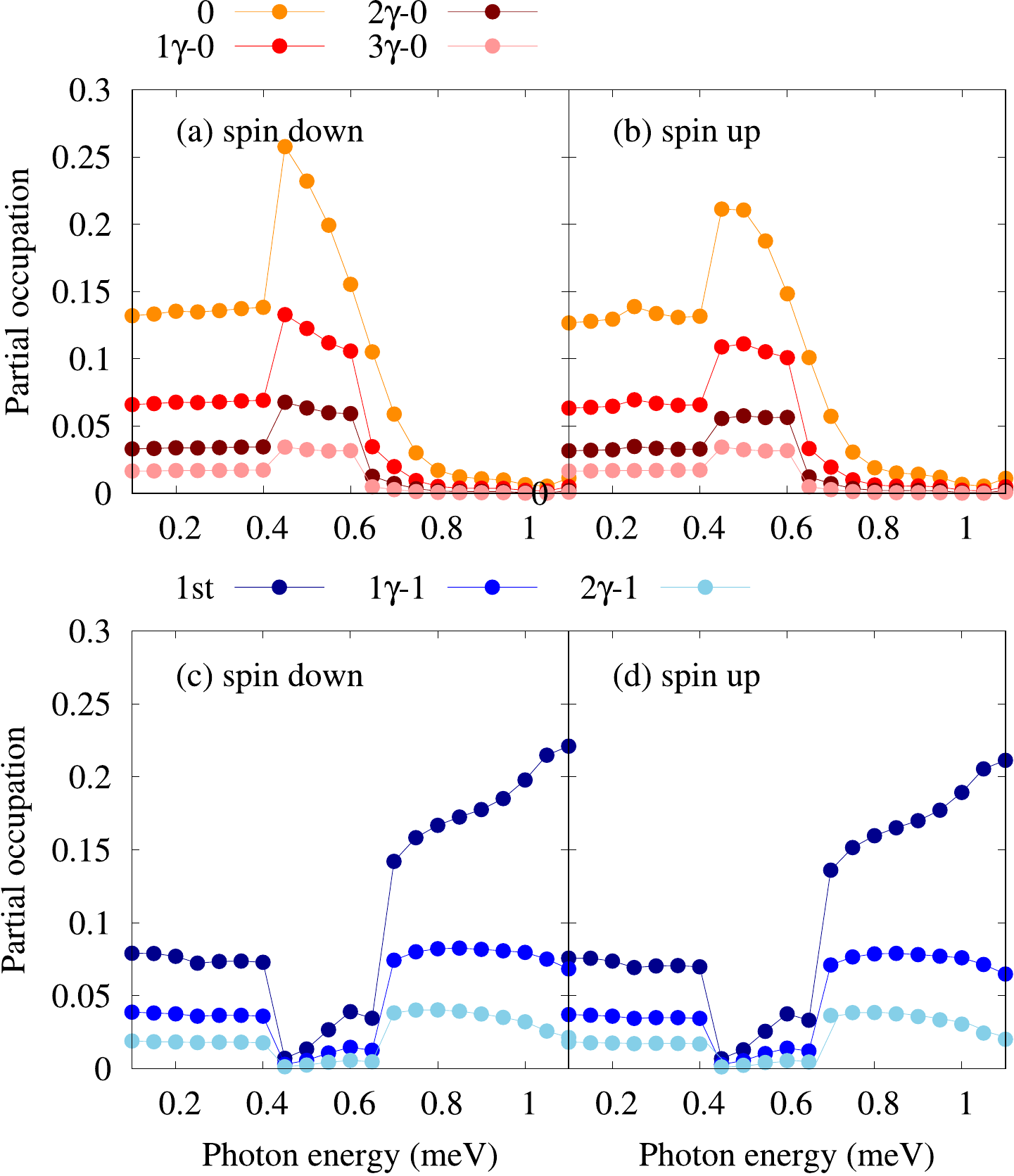}
	\caption{Partial occupation as a function of photon energy of the DQD system for the ground and its photon replica states with spin down (a) and spin up (b), and for first excited state and its photon replica states with spin down (c) and spin up (d) in the case of an $x$-polarized photon field. 0 is the one-electron ground-state (orange), $1\gamma\text{-}0$ (red), $2\gamma\text{-}0$(maroon), and $3\gamma\text{-}0$ (pink), respectively, refer to the first-, the second-, and the third-photon replica of the ground state. While $1^{\rm st}$ (dark blue) is the one-electron first-excited state, $1\gamma\text{-}1$ (blue) and $2\gamma\text{-}1$ (cyan)are the first- and the second-photon replica states of the first-excited state, respectively. The photon number initially in the cavity is $n_{\text{R}}=1$, $g_{\rm ph}=0.1$ meV and $\kappa=10^{-5}$ meV. The external magnetic field is $B=0.1$ T, $eV_{\rm g}=0.6$ meV, and $T_{L,R}=0.5$ K.}
	\label{fig04}
\end{figure}

The partial occupation for the ground and its photon replica states with spin down (a) and spin up (b), and for $1^{\rm st}$ and its photon replica states with spin down (c) and spin up (d) as a function of photon energy is presented in \fig{fig04}.
Almost a constant occupation can be seen at the ``low'' photon energy ranging from $0.1$ to $0.4$ meV, and the ``high" photon energy from $0.7$ to $1.1$~meV, whereas at the ``intermediate" photon energy from $0.4$ to $0.7$ meV a noticeable change occurs for all selected states with both spin down and spin up. The notable change in occupation comes from the Rabi resonances between the ground state and the first excited state with their replicas.
Consequently, the ground state and its replicas are further populated while the first-excited state and its replicas are depopulated in the ``intermediate" photon energy range for both spin components. The populated ground state and its replica at the intermediate photon energy or resonance region is caused by loss of the photons shown in \fig{fig03}, and the depopulation of the $1^{\rm st}$ and its replica comes from the gained photon content of these states.
The loss and gain of photons, and depopulation and the population of these states indicates that the system is under a stimulated emission-absorption process, a resonance.
It is interesting to see that the ground state, $0$, located below the bias window is highly populated while its photon content is almost zero and it is not in resonance with any other states for both directions of the photon polarization. The occupation of the $0$ state is caused by the stimulated emission-absorption process.

\begin{figure}[htb]
	\centering
	\includegraphics[width=0.45\textwidth]{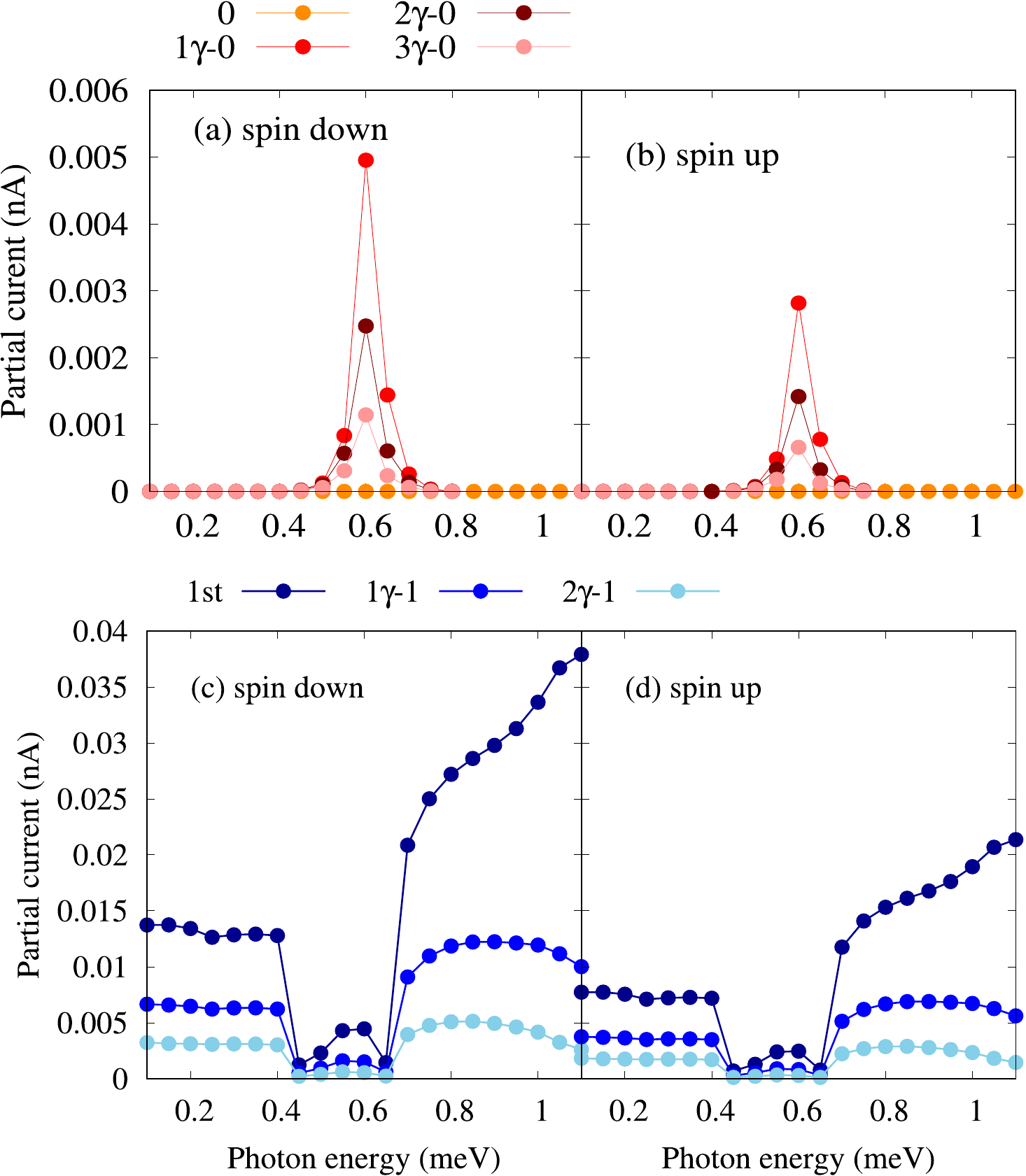}
	\caption{Partial current from the left lead to the DQD system as a function of photon energy for the ground and its photon replica states with spin down (a) and spin up (b), and for the first excited state and its photon replica states with spin down (c) and spin up (d) in the case of an $x$-polarized photon field. $0$ is the one-electron ground-state (orange), $1\gamma\text{-}0$ (red), $2\gamma\text{-}0$(maroon), and $3\gamma\text{-}0$ (pink), respectively, refer to the first-, the second-, and the third-photon replica of the ground state. While $1^{st}$ (dark blue) is the one-electron first-excited state, $1\gamma\text{-}1$ (blue) and $2\gamma\text{-}1$ (cyan) are the first- and the second-photon replica states of the first-excited state, respectively. The photon number in the cavity initially is $n_{\text{R}}=1$, $g_{\rm ph}=0.1$ meV and $\kappa=10^{-5}$ meV. The external magnetic field is $B=0.1$ T, $eV_{\rm g}=0.6$ meV, and $T_{L,R}=0.5$ K.}
	\label{fig05}
\end{figure}

In order to better understand the electron transport properties of the DQD system, we display
the partial currents into or out-off the system in \fig{fig05} for the ground state and its photon replicas of the
(a) spin down and (b) the spin up components, and for the first-exited state and its photon replicas
(c) spin down and (d) spin up components.
It is expected that the current of the photon replicas of the ground state is enhanced as their occupation are increased in the resonance region, the intermediate photon energy region. But the current of the ground state is zero as it is located below the bias window or far away from the bias window (see \fig{fig02}). In addition, the current of the first-excited state and its photon replicas is suppressed which is expected as their occupation are decreased in the resonance region.

\begin{figure}[htb]
	\centering
	\includegraphics[width=0.45\textwidth]{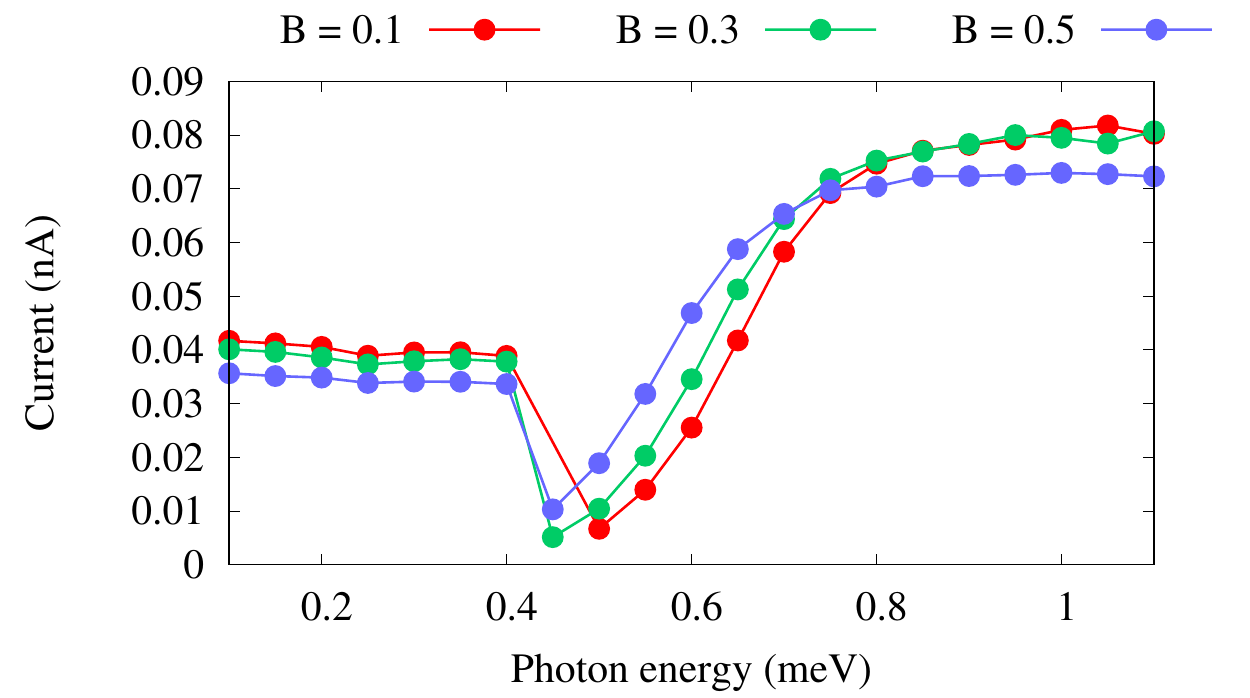}
	\caption{Total left current, the current from the left lead to the DQD system, as a function of the photon energy for $B = 0.1$ (red), $0.3$ (green), and $0.5$~T (blue) in the case of an $x$-polarized photon field. The photon number initially in the cavity is $n_{\text{R}}=1$, $g_{\rm ph}=0.1$ meV, $\kappa=10^{-5}$ meV, $eV_{\rm g}=0.6$ meV, and  $T_{L,R}=0.5$ K.}
	\label{fig06}
\end{figure}

The total left current, the current going from the left lead to the DQD, as a function of the photon energy is displayed in \fig{fig06}, where the influence of external magnetic field on the transport property of the DQD system is also shown. The characteristics of the total left current follow the behavior of the current of the first-excited state and its photon replica states, which is due to the presence of the first-excited state in the bias window. So, one can see a current dip in the intermediate photon energy in the total left current arsing from the multiple resonance states in this region.
The applied perpendicular magnetic field is slightly increased up two steps to $0.5$~T. The total left current is shown in \fig{fig06} for $B = 0.1$ (red), $0.3$ (green), and $0.5$ (blue).
The current dip is dislocated to a lower photon energy with a slightly higher value of current, when the magnetic field is increased in the resonance region or intermediate photon energy. In contrast, the current is decreased with increasing strength of the magnetic field in the low and high value of photon energy in the selected range of the photon energy.

\begin{figure}[htb]
	\centering
	\includegraphics[width=0.3\textwidth]{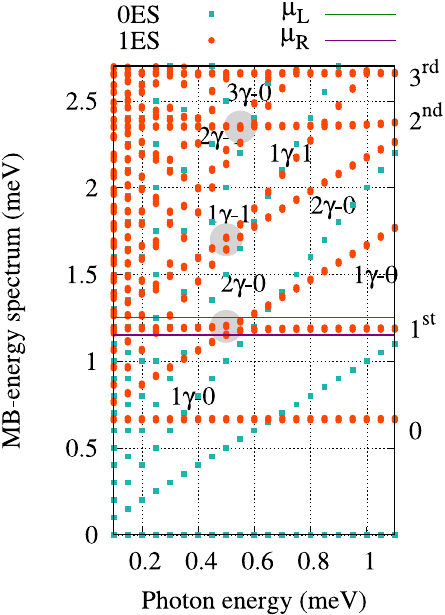}
	\caption{Many-Body (MB) energy spectrum of the DQD system coupled to a photon cavity as a function of the photon energy for the external magnetic field $B=0.5$ T and an $x$-polarized photon field, where 0ES (green squares) represent zero-electron states, and 1ES (red circles) are one-electron states. The green and purple lines are the chemical potential of the left lead, $\mu_L=1.25$ meV, and the right lead, $\mu_R=1.15$ meV, respectively. 0 is the one-electron ground-state energy, while $1^{\rm st}$, $2^{\rm nd}$, and $3^{\rm rd}$ are the one-electron first-, second-, and third-excited states respectively. $1\gamma\text{-}0$, $2\gamma\text{-}0$, and $3\gamma\text{-}0$, respectively, refer to one-, two-, and three-photon replicas of the ground state, while $1\gamma\text{-}1$ and $2\gamma\text{-}1$ are one- and two-photon replicas of the excited states, respectively. The initial photon number in the cavity is $n_{\text{R}}=1$, $g_{\rm ph}=0.1$ meV, $\kappa=10^{-5}$ meV, $eV_{\rm g}=0.6$ meV, and  $T_{L,R}=0.5$ K.}
	\label{fig07}
\end{figure}

The dislocation of the current dip at a higher value of the magnetic field, $B = 0.5$~T, can be explained by the energy spectrum and the photon content. The MB-energy spectrum of the DQD system for $B = 0.5$~T for an $x$-polarized photon field is shown in \fig{fig07}.

One can see that the anti-crossings in the MB-energy are shifted to a lower photon energy by increasing the applied magnetic field to $B=0.5$~T.
The main mechanism causing the anticrossing points to shift to lower energy as $B$ increases is the raising of the 1ES of the MB energy spectrum. This leads the anticrossing to occur at lower photon energy.
The degree of the dislocation of the anti-crossings is stronger at the lower part of MB-energy spectrum indicating that the external magnetic field is more effective at the lower part of the MB-energy spectrum.

\begin{figure}[htb]
	\centering
	\includegraphics[width=0.45\textwidth]{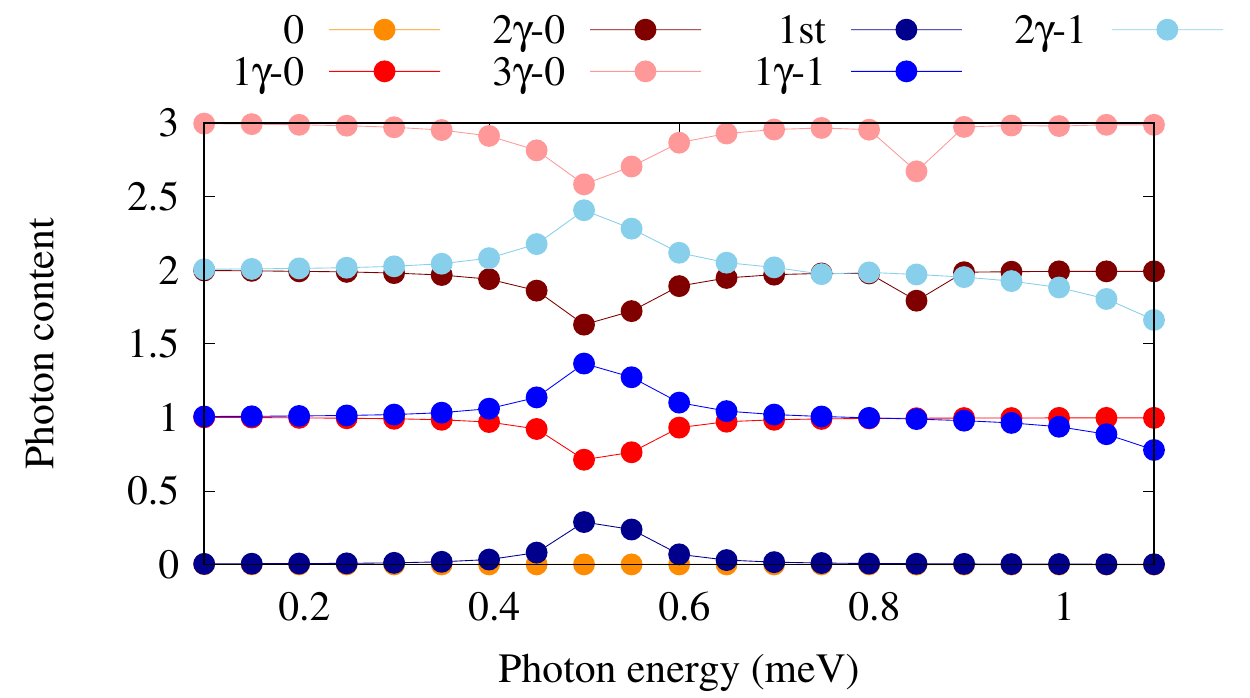}
	\caption{The photon content of the states as a function of photon energy for DQD system coupled to a cavity photon field, where the magnetic field is $B = 0.5$~T, and the photon content is independent of the directions of photon field. 0 is one-electron ground-state energy (orange), $1\gamma\text{-}0$ (red), $2\gamma\text{-}0$(maroon), and $3\gamma\text{-}0$ (pink), respectively, refer to first-, second-, and third-photon replica of ground state. While $1^{st}$ (dark blue) is one-electron first-excited state, $1\gamma\text{-}1$ (blue) and $2\gamma\text{-}1$ (cyan)are first- and second-photon replica states of first-excited state respectively. The initial photon number in the cavity is $n_{\text{R}}=1$, $g_{\rm ph}=0.1$ meV, $\kappa=10^{-5}$ meV, $eV_{\rm g}=0.6$ meV, and $T_{L,R}=0.5$ K.}
	\label{fig08}
\end{figure}

The strength of the anti-crossings can be better understood by providing the photon content of the states. The photon content in the case of $B = 0.5$~T is displayed in \fig{fig08}. The photon exchange position is dislocated to a lower photon energy at higher magnetic field, and one can clearly see that the photon exchange between the anti-crossing states is decreased by increasing magnetic field to $B = 0.5$~T. The photon exchange is getting weaker for the lower part of the energy spectrum such as $1^{\rm st}$ and $1\gamma-0$.
This is an indication that the Rabi resonance is weakened at the higher magnetic field. It thus leads to increased current in the resonance region and dislocates the current dip to a lower photon energy. We would expect to have a diminished current dip if the magnetic field is further increased. 
As the magnetic field is increased, the 1ESs of the MB-energy spectrum are shifted up in energy, which has been reported in \cite{PhysRevB.87.035314}.
We have found that the first photon replica forming the anticrossing in the bias window stays in the bias window till $B = 0.7$ T, and it is shifted up and goes slightly out of the bias window at higher magnetic field such as $B \ge 0.8$ T over the entire values of the selected range of the photon energy.
We therefore do not see any dip in the current, instead the current is almost zero.
The effects of the magnetic field on the resonance energy states and thus the electron transport have been reported in \cite{Rauf_Abdullah_2016}. It has been shown that the external magnetic field causes circular confinement of the charge density around the dot and diminishes the Rabi resonance effect.
Clearly, we are observing an interplay between the geometry of the DQD and the magnetic field.
The shapes of the charge density distribution of the states of the DQD system changes with
changing magnetic field.

We confirm that the current dip generated due to the multiple Rabi resonances
is further decreased with increasing $g_{\rm ph}$, as the multiple Rabi resonances are getting stronger with $g_{\rm ph}$. In contrast, if $\kappa$ is increased the current dip is enhanced and the minimum current in the dip vanishes. This is due to the fact that $\kappa$ weakens the multiple Rabi resonances in the system.

\section{Conclusions}\label{Conclusion}

A quantum master equation was utilized to investigate electron transport in a DQD-cavity system in a strong electron-photon coupling regime. By tuning the cavity parameters such as the photon frequency and photon polarization one can control the  electron motion in the DQD system.
We have seen a noticeable change in the current through the system when the photon frequency is tuned
reflected in a current dip arising from the multiple Rabi-resonances. The current dip is expected to be diminished when the strength of the external magnetic field applied to the DQD-cavity system is increased. In addition, we have realized that the direction of the photon polarization does not have much influence on electron transport which may be related to ``symmetric nature'' of the double quantum dots.

\section{Acknowledgment}
This work was financially supported by the University of Sulaimani and
the Research center of Komar University of Science and Technology.
The computations were performed on the IHPC facilities provided by the University of Iceland.

%\section{References}

%\bibliographystyle{elsarticle-num}
%\bibliography{Ref_2.bib}

\end{document}